\documentstyle[epsf]{mn}
\newif\ifAMStwofonts
\AMStwofontstrue

\ifoldfss
  \ifCUPmtlplainloaded \else
    \NewTextAlphabet{textbfit} {cmbxti10} {}
    \NewTextAlphabet{textbfss} {cmssbx10} {}
    \NewMathAlphabet{mathbfit} {cmbxti10} {} 
    \NewMathAlphabet{mathbfss} {cmssbx10} {} 
  \fi
  \ifAMStwofonts
    \ifCUPmtlplainloaded \else
      \NewSymbolFont{upmath} {eurm10}
      \NewSymbolFont{AMSa} {msam10}
      \NewMathSymbol{\upi}     {0}{upmath}{19}
      \NewMathSymbol{\umu}     {0}{upmath}{16}
      \NewMathSymbol{\upartial}{0}{upmath}{40}
      \NewMathSymbol{\leqslant}{3}{AMSa}{36}
      \NewMathSymbol{\geqslant}{3}{AMSa}{3E}

       \let\le=\leqslant
       \let\ge=\geqslant
    \fi
  \fi
\fi 

\ifnfssone
  \newmathalphabet{\mathit}
  \addtoversion{normal}{\mathit}{cmr}{m}{it}
  \addtoversion{bold}{\mathit}{cmr}{bx}{it}
  \newmathalphabet{\mathbfit} 
  \addtoversion{normal}{\mathbfit}{cmr}{bx}{it}
  \addtoversion{bold}{\mathbfit}{cmr}{bx}{it}
  \newmathalphabet{\mathbfss} 
  \addtoversion{normal}{\mathbfss}{cmss}{bx}{n}
  \addtoversion{bold}{\mathbfss}{cmss}{bx}{n}
  \ifAMStwofonts
    \ifCUPmtlplainloaded \else
      \UseAMStwoboldmath
      \makeatletter
      \new@mathgroup\upmath@group
      \define@mathgroup\mv@normal\upmath@group{eur}{m}{n}
      \define@mathgroup\mv@bold\upmath@group{eur}{b}{n}
      \edef\UPM{\hexnumber\upmath@group}
      \new@mathgroup\amsa@group
      \define@mathgroup\mv@normal\amsa@group{msa}{m}{n}
      \define@mathgroup\mv@bold\amsa@group{msa}{m}{n}
      \edef\AMSa{\hexnumber\amsa@group}
      \makeatother
      \mathchardef\upi="0\UPM19
      \mathchardef\umu="0\UPM16
      \mathchardef\upartial="0\UPM40
      \mathchardef\leqslant="3\AMSa36
      \mathchardef\geqslant="3\AMSa3E

       \let\le=\leqslant
       \let\ge=\geqslant
    \fi
  \fi
\fi 

\ifnfsstwo
  \DeclareMathAlphabet{\mathbfit}{OT1}{cmr}{bx}{it}
  \SetMathAlphabet\mathbfit{bold}{OT1}{cmr}{bx}{it}
  \DeclareMathAlphabet{\mathbfss}{OT1}{cmss}{bx}{n}
  \SetMathAlphabet\mathbfss{bold}{OT1}{cmss}{bx}{n}
  \ifAMStwofonts
    \ifCUPmtlplainloaded \else
      \DeclareSymbolFont{UPM}{U}{eur}{m}{n}
      \SetSymbolFont{UPM}{bold}{U}{eur}{b}{n}
      \DeclareSymbolFont{AMSa}{U}{msa}{m}{n}
      \DeclareMathSymbol{\upi}{0}{UPM}{"19}
      \DeclareMathSymbol{\umu}{0}{UPM}{"16}
      \DeclareMathSymbol{\upartial}{0}{UPM}{"40}
      \DeclareMathSymbol{\leqslant}{3}{AMSa}{"36}
      \DeclareMathSymbol{\geqslant}{3}{AMSa}{"3E}

       \let\le=\leqslant
       \let\ge=\geqslant
    \fi
  \fi
\fi 

\ifCUPmtlplainloaded \else
  \ifAMStwofonts \else 
    \def\upi{\pi}
    \def\umu{\mu}
    \def\upartial{\partial}
  \fi
\fi

\title{Wavelets Applied to CMB Maps: a Multiresolution Analysis for Denoising}
\author[] {J.L. Sanz$^{1}$, F. Arg\"ueso$^{2}$, L. Cay\'on$^{1}$, 
E. Mart\'\i nez-Gonz\'alez$^{1}$,R.B. Barreiro$^{1,3}$,
\newauthor
L. Toffolatti$^{4,5}$\\
1. Instituto de F\'\i sica de Cantabria, Fac. Ciencias, Av. los
	Castros s/n, 39005 Santander, Spain\\
2. Dpto. de Matem\'aticas, Universidad de Oviedo, c/ Calvo Sotelo s/n, 33007 Oviedo, Spain\\
3. Departamento de F\'\i sica Moderna, Universidad de Cantabria, 
	39005 Santander, Spain.\\
4. Dpto. de F\'\i sica, Universidad de Oviedo, c/ Calvo Sotelo s/n, 33007 Oviedo, Spain\\
5. Osservatorio Astronomico di Padova, vicolo dell'Osservatorio n5, 35122 Padova, Italy\\}

\date{\today}

\pagerange{\pageref{firstpage}--\pageref{lastpage}}
\pubyear{1998}

\begin{document}

\maketitle

\label{firstpage}

\begin{abstract}

Analysis and denoising of Cosmic Microwave Background 
(CMB) maps are performed using 
wavelet multiresolution techniques. The method is tested on 
$12^{\circ}.8\times 12^{\circ}.8$ maps with resolution resembling
the experimental one expected for future high resolution space 
observations. 
Semianalytic formulae of the variance of wavelet coefficients 
are given for the Haar and Mexican Hat wavelet bases. Results are
presented for the standard Cold Dark Matter (CDM) model. 
Denoising of simulated maps is carried out by removal of 
wavelet coefficients dominated by instrumental noise. CMB maps with a 
signal-to-noise, $S/N \sim 1$, 
are denoised with an error improvement factor between 3 and 5. 
Moreover we have also tested how well
the CMB temperature power spectrum is recovered after denoising. We are
able to reconstruct the $C_{\ell}$'s up to $l\sim 1500$ with errors 
always below $20\% $ in cases with $S/N \ge 1$.

\end{abstract}

\begin{keywords}
cosmology: CMB -- data analysis
\end{keywords}

\section{Introduction}

	Future CMB space experiments will provide very detailed all-sky maps 
of CMB temperature anisotropies; NASA MAP Mission 
(Bennett et al. 1996)
and the ESA Planck Mission (Mandolesi et al. 1998;
Puget et al. 1998).  The high sensitivity of these experiments will result 
in unique data to constrain fundamental cosmological parameters. Moreover,
future CMB maps will allow to distinguish between competing theories of
structure formation in the early universe and will provide very 
fruitful data on astrophysical foregrounds.

The cosmological signal in CMB maps is hampered by instrumental noise and
by foreground emissions. Therefore, a necessary step in analysing CMB maps 
is
to separate the foreground emissions from the CMB signal. Several linear
and non-linear
methods have already been tested on simulated data (Bouchet,
Gispert \& Puget 1996; Tegmark \& Efstathiou 1996; Hobson et al. 1998a,b).  
An alternative method can be one based on wavelets. 
Wavelets are known 
to be very efficient in dealing with problems of data compression and 
denoising. 
Development of wavelet techniques applied to signal processing 
has been very fast in the last ten years (see Jawerth \& 
Sweldens 1994 for an overview). These techniques have already been applied 
to a variety of astrophysical problems. For example, regarding
cosmology, Slezak, 
de Lapparent \& Bijaoui (1993) have applied wavelet analysis to the detection of
structures in the CfA redshift survey. They have also been introduced to 
study the Gaussian character of CMB maps (Pando et al. 1998, Hobson et al. 1998). A study using spherical Haar wavelets to denoise CMB maps has just appeared
(Tenorio et al. 1999). 

We consider small patches of the sky
where a flat 2-D approach is valid.
We apply wavelet multiresolution techniques, known to be computationally very 
fast taking only $O(N)$ operations to reconstruct an image of $N$ pixels. 
In the 2-D flat 
wavelet analysis a 
single scale and two translations are usually introduced, where the basis
is generated 
by 4 tensor products of wavelets and scaling functions. Therefore,
three {\it detail} images plus an {\it approximation} image appear
at each level of resolution. Different wavelet bases are characterized
by their location in space. The bases considered in this work are
Mexican Hat, Haar and Daubechies. The first one is the most localized
though, as opposed to the other two, it does not have a compact support.
As a first approach to the application of these techniques to CMB data
we only consider maps with cosmological signal plus instrumental 
Gaussian noise. 
CMB experiments are contaminated by galactic (dust, 
free-free and synchrotron emission) and extragalactic foregrounds (infrared 
and radio galaxies, S-Z effects from clusters,...) in addition to noise. 
Obviously in view of this, the contents of this paper only cover
a small fraction of the work to be done. 
Our aim is to 
shed light on the wavelet characterization of the different components,
the late phase being 
to build up a wavelet-based framework to disentagle all of them.
In this
line, a Bayessian method (maybe incorporating entropy or other constraints) 
dealing with wavelet components at different scales and integrating 
the different channels will be the final goal. 
Knowing the efficiency of wavelets in removing noise as shown by many
works in other fields, the first step to take in the application
of wavelet techniques to the CMB is to study the denoising of 
temperature maps.

The outline of the paper is as follows. A theoretical 
continuous wavelet analyis of CMB data is presented in Section 2. General
semianalytical formulae are given for the variance of the {\it detail}
wavelet coefficients as a function of the temperature power
spectrum $C_{\ell}$. Section 3 introduces the discrete wavelet technique that
is applied to denoising of simulated CMB maps in Section 4. Discussion and
conclusions are presented in Section 5.

\setcounter{figure}{0}
\begin{figure}
 \epsfxsize=3in
 \epsffile{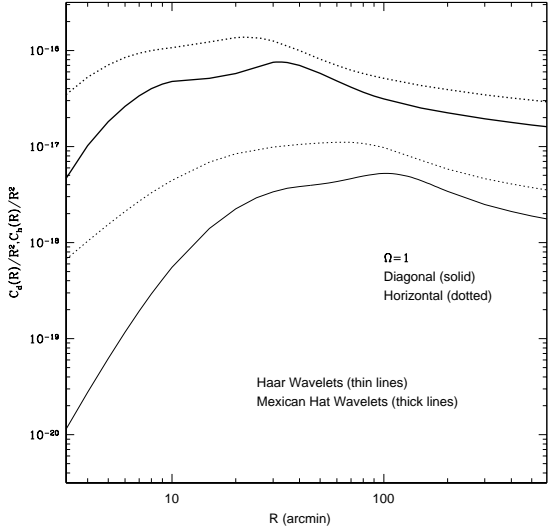}
 \caption{
Variance of diagonal (solid lines) and horizontal/vertical 
(dotted lines) {\it detail} 
wavelet coefficients $C_d,C_h$ versus scale $R$. A standard CDM cosmological
model is assumed. Thin lines outline the result obtained for the Haar wavelet
system and the thick lines correspond to the Mexican Hat wavelet basis.
}
 \label{f1}
\end{figure}

\section{Continuous Wavelet Analysis}

\subsection{One-dimensional Transform}

The Fourier transform is a powerful tool in many areas but in dealing with
local behaviour shows a tremendous inefficiency. For instance, a large 
number of complex exponentials must be combined in order to produce a spike.
The wavelet transform solves this problem, introducing a good space-frequency
localization. It is conceptually simple and it constitutes a fast algorithm.   
Let $\psi (x)$ be a one-dimensional function satisfying the following
conditions: a) $\int_{-\infty}^{\infty}dx\,\psi (x) = 0$, b)
$\int_{-\infty}^{\infty}dx\,{\psi}^2 (x) = 1$ and 
c) $C_{\psi}\equiv \int_{-\infty}^{\infty}dk\,|k|^{-1}\psi (k) < \infty $, 
where $\psi (k)$ is
the Fourier transform of $\psi (x)$. So, according to condition a), the wavelet
must have oscillations. Condition b) is a normalization and c) represents an
admissibility condition in order to reconstruct a function $f(x)$ with the
basis $\psi$ (see equation (2) for such a synthesis). 

\noindent We define the {\it analyzing} wavelet as 
$\Psi (x; R, b) \equiv R^{-1/2}\psi (\frac{x - b}{R})$, dependent on two
parameters: dilation ($R$) and translation ($b$). It operates as a mathematical
microscope of magnification $R^{-1}$ at the space point $b$. The wavelet 
coefficients associated to a one-dimensional function $f(x)$ are: 

\begin{equation}
w(R, b) = \int dx\, f(x)\Psi (x; R, b)\; .
\end{equation}

\noindent It is clear from the above definition that 
such coefficients represent the
analyzing wavelet at $x_o$ for a delta distribution peaked at this point, i.e.
for $f(x) = \delta (x - x_o)$. For $R = 1$, $w(R, b)$ is the convolution of the
function $f$ with the analyzing wavelet $\psi$.
  
\noindent The reconstruction of the function $f$ can be achieved in the form

\begin{equation}
f(x) = (2\pi C_{\psi})^{-1}\int \int dR\,db\, R^{-2}w(R, b)\,
\Psi (x; R, b)\; .
\end{equation}

\noindent Examples of wavelet functions are: i) Haar, $\psi = 1 (-1)$ for 
$0 < x < 1/2$ $(1/2 <x < 1)$, ii) Mexican hat, 
$\psi = \frac{2}{{(9\pi)}^{1/4}}(1 - x^2)e^{-x^2/2}$.


\setcounter{figure}{1}
\begin{figure*}
 \epsfxsize=6in
 \epsffile{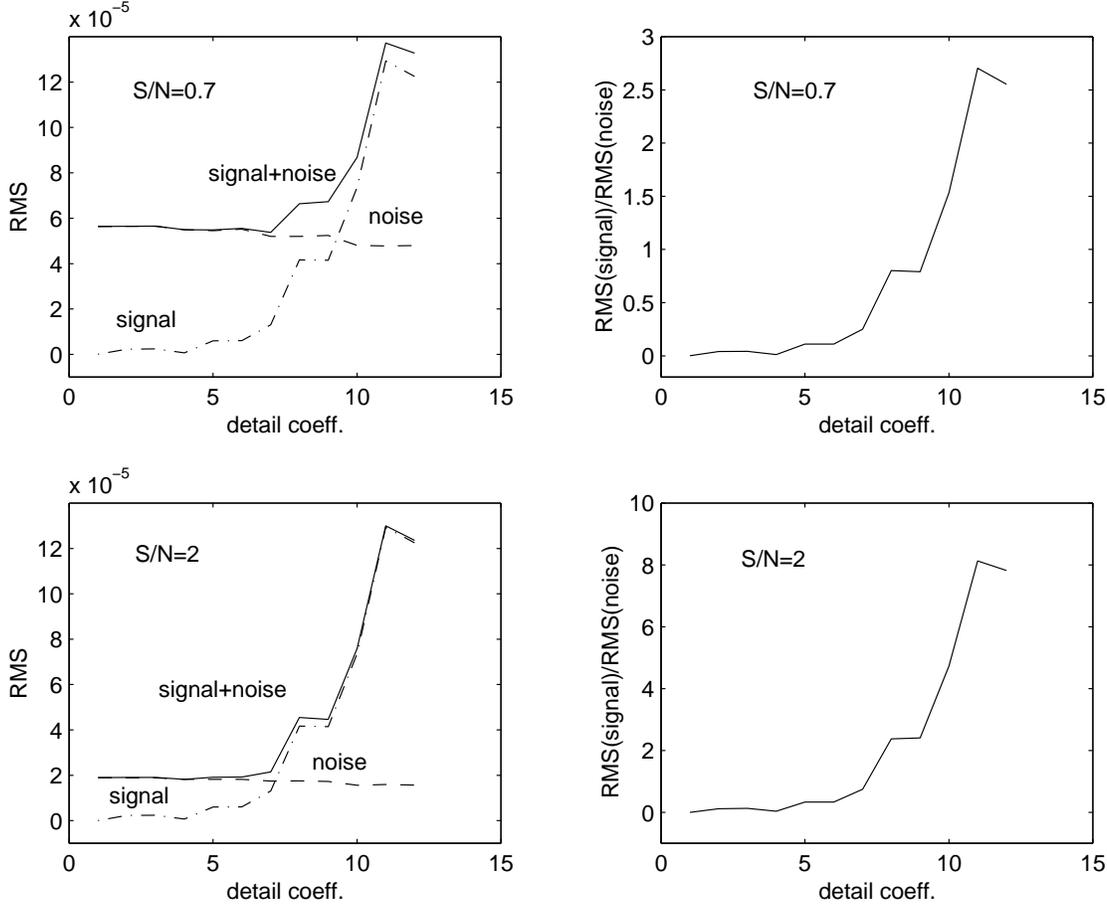}
 \caption{
rms deviation of wavelet {\it detail} coefficients 
obtained from CMB maps (signal maps, dashed-dotted lines), CMB plus noise maps
(signal plus noise maps, solid lines) and pure noise (dashed lines) are
presented in left panels. Right panels show the ratio of the rms deviation
of the {\it detail} coefficients from
signal maps divided by the rms deviation of the {\it detail} coefficients 
from noise maps. Top panels correspond to 
simulated maps with
$S/N=0.7$; bottom panels correspond to 
$S/N=2$.
}
 \label{f2}
\end{figure*}

\subsection{Two-dimensional Transform}

Regarding the two-dimensional case, we introduce a one-dimensional {\it
scaling} function $\phi$ normalized in the form:
$\int_{-\infty}^{\infty}dx\,\phi (x) = 1$. 
Examples of scaling functions are: i) Haar, $\phi =1$ $(0)$ for $0<x<1$
$(x<0,x>1)$, ii) Mexican Hat, $\phi =\frac{2}{{(9\pi)}^{1/4}}e^{-x^2/2}$.
The analyzing {\it scaling}
$\Phi (x; R, b) \equiv R^{-1/2}\phi (\frac{x - b}{R})
$, allows to define 
{\it details} of an image, $f(\vec{x})$, with respect to the tensor products

\begin{equation}
{\Gamma }_d(\vec{x}; R, \vec{b}) \equiv \Psi (x_1; R, b_1)\Psi 
(x_2; R, b_2)\; ,
\end{equation}
\begin{equation}
{\Gamma }_h(\vec{x}; R, \vec{b}) \equiv \Phi (x_1; R, b_1)\Psi (x_2; R, b_2)\; ,
\end{equation}
\begin{equation}
{\Gamma }_v(\vec{x}; R, \vec{b}) \equiv \Psi (x_1; R, b_1)\Phi (x_2; R, b_2)\; .
\end{equation}

\noindent The {\it diagonal, horizontal} and {\it vertical} wavelet 
coefficients 
are defined by ($\alpha \equiv d, h, v$)

\begin{equation}
w_{\alpha}(R, \vec{b}) = \int d\vec{x}\,f(\vec{x})\,{\Gamma}_{\alpha}
(\vec{x}; R, \vec{b})\; .
\end{equation} 
\noindent Scaling functions act as low-pass filters whereas wavelet functions
single out one scale.
Therefore, {\it detail} coefficients provide local 
information about symmetrical
(diagonal) and elongated/filamentary structure (vertical and horizontal).

Let us now assume an homogeneous an isotropic random field $f(\vec{x})$, i.e.
the correlation function 
$C(r) \equiv <f(\vec{x})f(\vec{x} + \vec{r})>$, $r
\equiv |\vec{r}|$, where
$<>$ denotes an average value over realizations of the field. The Fourier
transform of the field $f(\vec{k})$ satisfies $<f(\vec{k})
f({\vec{k}}^{\prime})> =
P(k){\delta}^2(\vec{k} - \vec{k^{\prime}})$, where $k \equiv |\vec{k}|$ and 
$P(k)$ is the power spectrum (the Fourier transform of $C(r)$). In this case we
can calculate the correlation and variance of the wavelet coefficients:
$C_{\alpha}(r; R) \equiv <w_{\alpha}(R, \vec{b})w_{\alpha}(R, \vec{b} + 
\vec{r})>, {\sigma}^2_{\alpha}(R) \equiv C_{\alpha}(0; R)$ and we find the
following equations

\begin{eqnarray}
C(0)\equiv {\sigma}^2 = C_{\Gamma_{\alpha}}^{-1}\int
dR\,R^{-3}{\sigma}^2_{\alpha}(R)\; ,\nonumber
\\
C_{\Gamma_{\alpha}} \equiv {(2\pi)}^2\int
d\vec{k}\,k^{-2}|{{\tilde{\Gamma}}_{\alpha}|}^2(\vec k)\; ,
\end{eqnarray}

\noindent where ${\tilde{{\Gamma}}}_{\alpha}(\vec{k})$ is the Fourier 
transform of $R{\Gamma}_{\alpha}$. 

On the other hand, we calculate the Fourier transform of the wavelet 
coefficients
$w_{\alpha}(R, \vec{b})$ with respect to the $\vec{b}$ parameters:

\begin{equation}
<w_{\alpha}(R, \vec{k})w_{{\alpha}^{\prime}}(R^{\prime}, {\vec{k}}^{\prime})> 
= w_{\alpha {\alpha}^{\prime}}(R, R^{\prime};\vec {k}){\delta}^2(\vec{k} - 
\vec{k^{\prime}})\; ,
\end{equation}
\begin{equation}
w_{\alpha {\alpha}^{\prime}} = {(2\pi)}^2 RR^{\prime}P(k)
{\tilde{\Gamma}}_{\alpha}^*(R\vec{k})
{\tilde{\Gamma}}_{{\alpha}^{\prime}}(R^{\prime}{\vec{k}})\; ,
\end{equation} 

\noindent that allows us to get the {\it detail} wavelet variances as

\begin{equation}
{\sigma}^2_{\alpha}(R) = \int
d\vec{k}\,P(kR^{-1}){|{\tilde{\Gamma}}_{\alpha}(\vec{k})|}^2\; .
\end{equation}

The diagonal variance corresponds to the tensor product of two one-dimensional
wavelets. If $|\psi (k)|^2$ is a function strongly peaked near $k\simeq 1$ 
then ${\sigma}^2_d(R) \simeq P(k\simeq R^{-1})$, taking into account the
normalization of the wavelet function, that allows an estimation of 
the power spectrum in terms of the diagonal component. This is what happens 
for the Mexican hat:
$|\psi (k)|^2 \propto k^4e^{-k^2}$, with a maximum at $k = 2^{-1/2}$, whereas
the Haar wavelet is not localized in Fourier space: 
$|\psi (k)|^2 \propto (k/4)^{-2}sin^4(k/4)$. We can also deduce that
$C_h=C_v$ and
${\sigma}^2_h = {\sigma}^2_v$ taking into account the symmetry of
the equations. Moreover, the temperature power spectrum $P(k)$ can be 
obtained from the {\it detail} wavelet power spectrum 
$w_{\alpha \alpha }(R,R;\vec k)$ as follows
\begin{eqnarray}
P(k)={{1}\over {C_{\Gamma _{\alpha}}}} \int {{dR}\over {R^{-3}}} \int 
d\theta \, w_{\alpha \alpha }(R,R;k\vec n)\; , \nonumber\\
\vec n=(\cos\theta , \sin\theta )\; .
\end{eqnarray}

For the Haar and Mexican wavelets we can calculate:

\begin{equation}
HAAR: {|{\tilde{\Gamma}}_d|}^2 = \frac{1}{{(2\pi)}^2}
{(\frac{k_1k_2}{4})}^{-2}[\sin \frac{k_1}{4} \sin \frac{k_2}{4}]^4\; ,
\end{equation}
\begin{equation}
{|{\tilde{\Gamma}}_h|}^2 = \frac{1}{{(2\pi)}^2}
{(\frac{k_1k_2}{4})}^{-2}[\sin \frac{k_2}{4}]^4 \frac{1}{4}{\sin}^2 k_1\; ,
\end{equation}

\begin{eqnarray}
MEXHAT: {|{\tilde{\Gamma}}_d|}^2 = \frac{16}{9\pi}{(k_1k_2)}^4e^{-k^2},\nonumber
\\
{|{\tilde{\Gamma}}_h|}^2 = \frac{4}{3\pi}{k_2}^2e^{-k^2},
\end{eqnarray}

\noindent where $k^2 = k_1^2 + k_2^2$ and $\tilde{\Gamma}_v$ can be obtained
from $\tilde{\Gamma}_h$, swapping $k_1$ and $k_2$.
The variance of the {\it detail} wavelet coefficients for the Haar and 
Mexican Hat 
systems, assuming the standard CDM model, is presented in Figure 1. 
As one can see the acoustic peaks can be clearly noticed,
being more pronounced for the Mexican Hat basis. This last result is a 
consequence of being a more localized wavelet system. For a more detailed
discussion see Sanz et al. 1998, 1999.


\section{Discrete Wavelet Analysis}

\subsection{One-dimensional Multiresolution Analysis}

An orthonormal basis of $L^2(\Re )$ can be constructed from a 
wavelet $\psi$ through dyadic dilations $j$  and translations $k$
\begin{equation}
\psi_{j,k}(x)=2^{j/2}\psi(2^jx-k)\; .
\end{equation}
\noindent In addition, a scaling function $\phi$
can be defined associated to the {\it mother} wavelet $\psi $. Such a 
function gives rise to the so called multiresolution analysis. 
A multiresolution analysis of $L^2(\Re )$ is defined as a sequence of
closed subspaces $V_j$ of $L^2(\Re )$, $j\in Z$. Properties can be 
seen in Ogden (1997).  

Subspaces $V_j$ are generated by dyadic dilations and translations
of the scaling function $\phi $ (this function forms an orthonormal
basis of $V_o$, $\{ \phi_{o,k}(x)=\phi(x-k)\}$). Moreover each $V_j$ can be expressed as the orthogonal sum $V_j=V_{j-1}\oplus W_{j-1}$,
where $W_{j-1}$ is created from wavelets $\psi_{j-1,k}$.
Taking into account the properties of the scaling function,
together with this last expression, we can construct {\it approximations} at 
increasing levels of resolution. These {\it approximations} are linear combinations
of dilations and translations of a scaling function $\phi $. The difference
between two consecutive {\it approximations}, i.e. the  {\it detail} 
at the corresponding resolution level, is given by a linear combination 
of dilations and 
translations of a wavelet function $\psi $.

\setcounter{figure}{2}
\begin{figure*}
 \epsfxsize=6in
 \epsffile{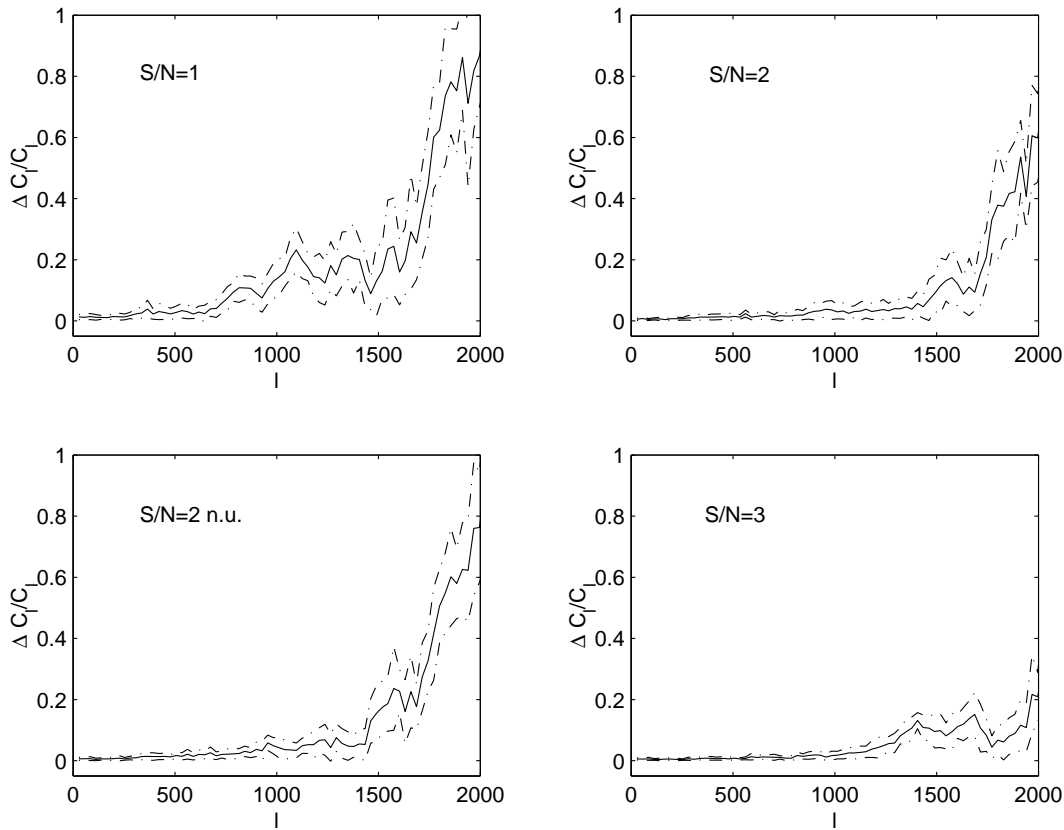}
 \caption{
Mean value (solid line) and 
$1\sigma $ error (dashed-dotted lines) of the
absolute value of the relative errors, $\Delta C_{\ell}/C_{\ell}$.
Top-left panel corresponds to $S/N=1.0$,
top-right to $S/N=2$, bottom-left to $S/N=2$ with non-uniform noise and
bottom-right to $S/N=3$.
}
 \label{f3}
\end{figure*}

\subsection{Two-dimensional Multiresolution Analysis}

The analysis performed in this work assumes equal dilations in the
2 dimensions involved. 
At a fixed level of resolution, subspaces in a 
2-D multiresolution analysis are the tensor products of the
corresponding one-dimensional ones 
${\bf V}_{j+1}=V_{j+1}\otimes V_{j+1}$.The 2-D basis is therefore built 
by the product of two scaling functions ({\it approximation}), the product of
wavelet and scaling functions (horizontal and
vertical {\it details}) and the product of two wavelets (diagonal {\it details}):
\begin{equation}
{\bf V}_{j+1}=(V_j\oplus W_j)\otimes (V_j\oplus W_j)
\end{equation}
\begin{equation}
~~~~~~~~~~~~~=(V_j\otimes V_j)\oplus [(V_j\otimes W_j)\oplus (W_j\otimes V_j)
\oplus (W_j\otimes W_j)]\; .
\end{equation}
\noindent Horizontal, vertical and diagonal {\it detail} coefficients 
represent the variations
in these directions relative to a weighted average at a lower resolution 
level (given by the {\it approximation} coefficients).  

A discrete orthonormal basis, ${\Gamma}_{\alpha}(\vec{x}; j, \vec{k})$, can
be defined by setting
$R = 2^{-j}$ and $\vec{b} = 2^{-j}\vec{k}$ in equations (3-5), then
$({\Gamma}_{\alpha}(\vec{x}; j, \vec{k}){\Gamma}_{{\alpha}^{\prime}}(\vec{x};
j^{\prime}, {\vec{k}}^{\prime})) = {\delta}_{\alpha
{\alpha}^{\prime}}{\delta}_{jj^{\prime}}{\delta}_{\vec{k}{\vec{k}}^{\prime}}$,
where $()$ denotes the scalar product in $L^2(\Re ^2)$. If we define the 
discrete 
wavelet coefficients associated to any {\it detail} by the equation (6) 

\begin{equation}
w_{\alpha}(j, \vec{k}) = \int d\vec{x}\,f(\vec{x})\,{\Gamma}_{\alpha}
(\vec{x}; j, \vec{k})\; ,
\end{equation}

\noindent we can thus reconstruct the image with all the {\it details}

\begin{equation}
f(\vec{x}) = {\sum}_{\alpha , j, \vec{k}}w_{\alpha}(j, \vec{k})
{\Gamma}_{\alpha}(\vec{x}; j, \vec{k})\; .
\end{equation}

\noindent In particular, we get the following expression for the second-order
moment of the image

\begin{equation}
(f^2(\vec{x})) = {\sum}_{\alpha , j, \vec{k}}w^2_{\alpha}(j, \vec{k})\; ,
\end{equation}

\noindent that expresses how the energy of the field is distributed locally at 
any scale and {\it detail}. 

For a finite image, $R_{max}\times R_{max}$, in order
to reconstruct it we must add to equation (19)
an approximation $w_a(\vec{k}){\Gamma}_a(\vec{x}; \vec{k})$ with
${\Gamma}_a(\vec{x}; \vec{k})\equiv \Phi (x_1; R_{max},k_1)$
$\Phi (x_2; R_{max},k_2)$ and $w_a(\vec{k}) \equiv \int
d\vec{x}\,f(\vec{x})\,{\Gamma}_a(\vec{x},\vec k)$, representing the field at the lower
resolution. If $f(\vec x)$ represents the temperature fluctuation field then
the variance is given by $<(\Delta T/T)^2>=((\Delta T/T)^2)/N_p$, being 
$N_p$ the number of pixels. 

The orthonormal basis that we are going to use are the standard Daubechies $N$
(Haar corresponds to $N = 1$), that has been extensively used in the
literature because of their special properties: they are defined in a compact 
support, have increasing regularity with $N$ and vanishing moments up to order 
$N - 1$ (Daubechies 1988). On the contrary, the Mexican Hat wavelet is not
defined in a compact support and it is not appropriate for this multiresolution
analysis. 

For discrete wavelet analysis of the CMB maps we use 
the Matlab Wavelet Toolbox (Misiti et al. 1996).
This toolbox is an extensive collection of 
programs for analyzing,
denoising and compressing signals and 2-D images.
Discrete Wavelet decomposition is performed as described above
to obtain the {\it approximation}
and {\it detail} coefficients of the 2-D CMB maps at several levels.


\setcounter{figure}{3}
\begin{figure*}
 \epsfxsize=6in
 \epsffile{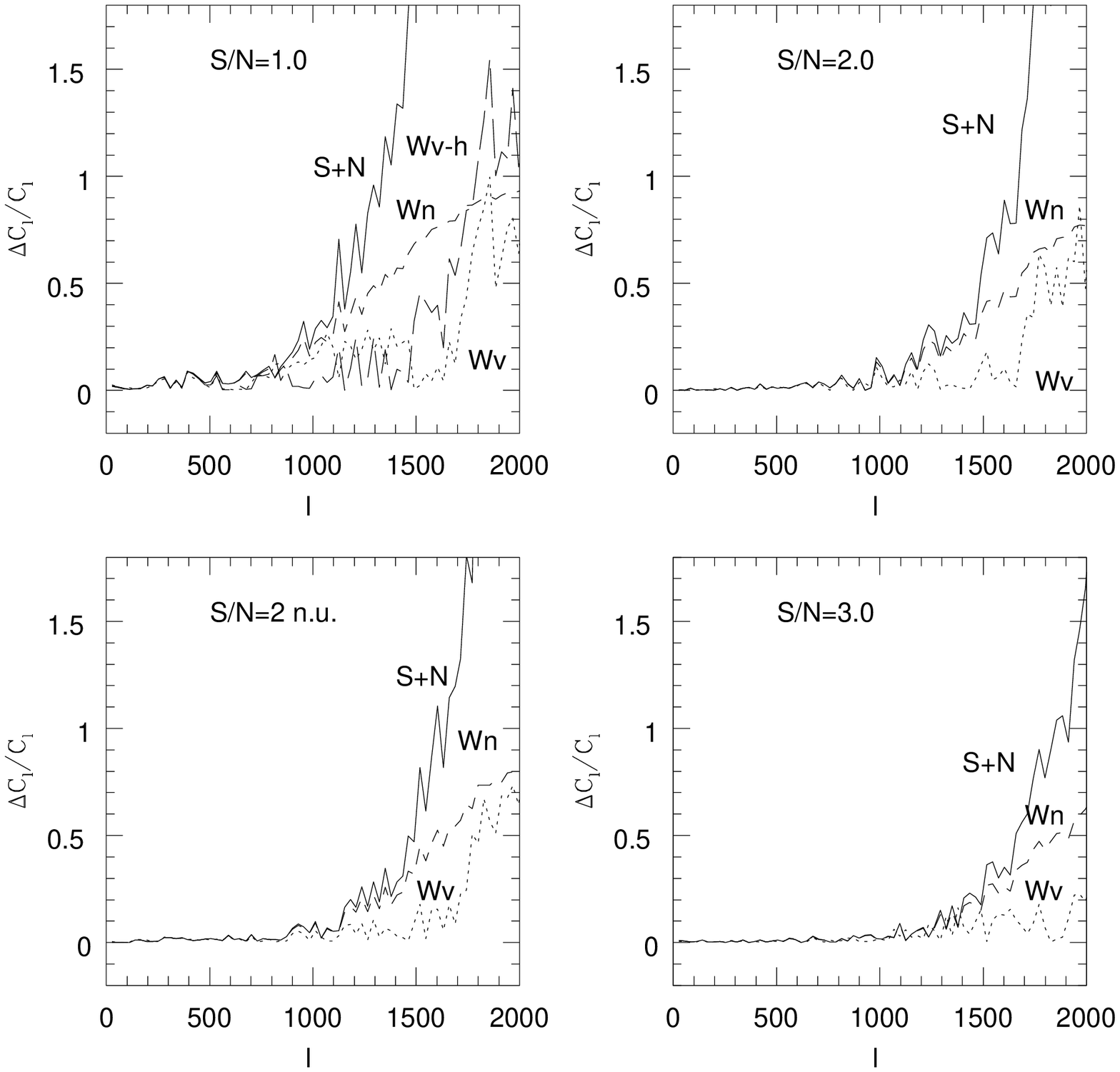}
 \caption{
Absolute value of the relative errors, $\Delta C_{\ell}/C_{\ell}$, of the 
CMB power spectrum 
obtained from signal-plus-noise maps (solid lines), wavelet denoised maps
(short dashed lines) and Wiener denoised maps (dashed lines). 
Top-left panel corresponds to $S/N=1$ (wavelet denoised maps removing
all coefficients at levels 1, 2, 3d and 3h is included 
as long dashed lines),
top-right to $S/N=2$, bottom-left to $S/N=2$ with non-uniform noise 
and
bottom-right to $S/N=3$. 
}
 \label{f4}
\end{figure*}

\section{Denoising of CMB Maps}

Future CMB space experiments will provide maps with resolution 
scales of few arcminutes.
In this work we analyze simulated maps of
$12.8\times 12.8$ square degrees with pixel size of $1.5$ arcmin.
Simulations are made assuming the standard CDM, $\Omega =1$ and $H_o=50$ km/sec/Mpc. 
The maps
are filtered with a $4'.5$ FWHM Gaussian beam to approximately 
reproduce the filtering scale of
the High Frequency Channels of the Planck Mission. Simulated maps 
have a rms signal of $\Delta T/T=3.7\times 10^{-5}$. Gaussian noise is added to these
maps at different $S/N$ levels between $0.7$ and 3. A non-uniform noise 
is also considered to account for the non-uniform sampling introduced in
satellite observations. As an extreme case we have assigned the
signal-to-noise at each pixel from a truncated (at the $2\sigma $ level) Gaussian distribution with 
a mean value of 2 and a dispersion of $0.5$.
We use the set of Matlab Wavelet 2-D programs with the corresponding 
graphical interface to analyze and denoise those maps. Suitable bases 
of wavelets are studied. Daubechies 4 wavelets are the ones used in this
analysis. No significant changes are observed when the analysis is carried
out using other higher order Daubechies bases. On the other hand, the Haar 
system is not so efficient for denoising CMB maps since it produces 
reconstruction errors much larger than using high order Daubechies systems.

\setcounter{figure}{4}
\begin{figure*}
 \epsfxsize=6in
 \epsffile{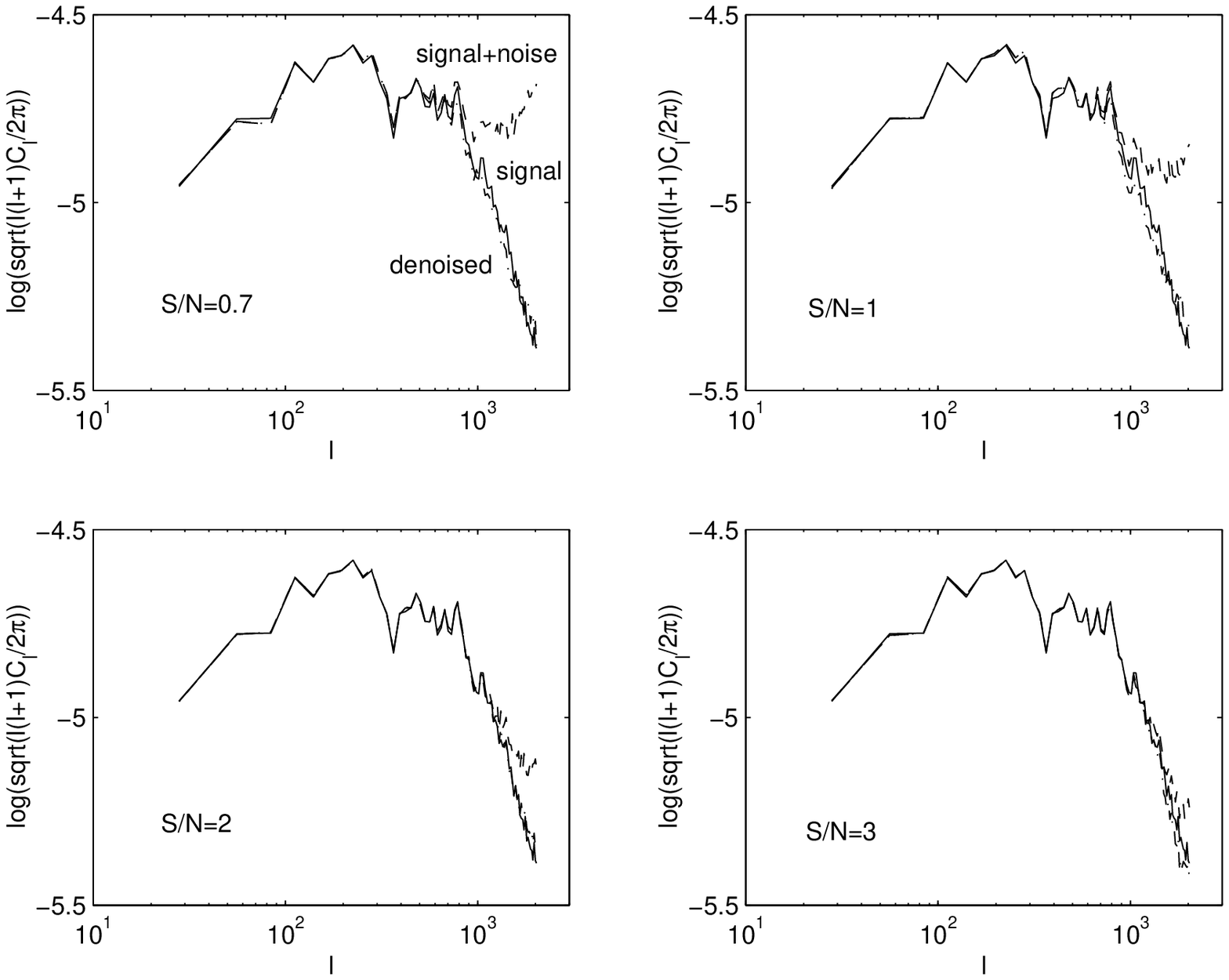}
 \caption{
Power spectrum obtained from signal-plus-noise maps
(dashed lines), signal maps (solid lines) and denoised maps (dashed-dotted 
lines). 
Top-left $S/N=0.7$, top-right $S/N=1$, bottom-left $S/N=2$ and
bottom-right $S/N=3$.
}
 \label{f5}
\end{figure*}

First of all, three wavelet decompositions are performed 
obtaining wavelet coefficients 
corresponding to the CMB original map, to the signal plus noise map and
to the pure noise map. Decompositions are carried out up to the fourth 
resolution level. 
Denoising of the
signal plus noise maps is based on subtraction of certain sets
of coefficients affected by noise. White noise is the most common
in CMB experiments. The dispersion of wavelet coefficients of that type
of noise is constant as can be seen from equation 10. On the  
contrary CMB detail wavelet dispersions go to zero as $R$ goes to zero. 
Therefore first level wavelet coefficients are dominated by noise
and then, for a given signal plus noise
map, it is possible to know the noise and consequently the CMB wavelet
coefficient dispersions at all levels. 
CMB maps produced by typical experiments with a ratio between
antenna and pixel size of $\approx 3$ will have wavelet coefficients 
containing the relevant information on the signal at level 3 and above. 
As shown below, level 3 is the critical one to perform denoising as 
the noise can still be at a level comparable to the signal.
Figure 2 shows
rms deviations and corresponding ratios 
for two simulations with $S/N=0.7$ and $S/N=2$. {\it Detail} coefficient numbering 
corresponds to the three directions diagonal, vertical and horizontal at
the three consecutive levels, i.e., numbers 1,2,3 correspond to
diagonal, vertical and horizontal cofficients at the first resolution level,
4,5,6 to the second level coefficients in the same order and 7,8,9,10,11,12 
to levels 3 and 4 respectively. As it can be seen, the first two levels are
entirely dominated by noise as pointed out before.
Therefore, all these coefficients can be 
removed to reconstruct a denoised map. This is equivalent to using a 
hard thresholding assuming a threshold above all these coefficients.
On the other side, level 4 is completely
dominated by the CMB signal and is left untouched. 
Ratios between rms deviations of the signal and
noise maps at the third resolution level are not always clearly dominated
by noise or signal. 
Ratios of $\approx 1$ are treated with a soft
thresholding technique (in practice we consider ratios in the range
$0.3-1.5$ though changes in this interval do not significantly 
affect results).  
Soft thresholding consists of removing all coefficients
with absolute values smaller than the threshold defined in terms of
the noise dispersion ($\sigma_n$). Coefficients with absolute values 
above the defined threshold are rescaled by subtracting the threshold to 
the positive ones and adding it to the negative ones. 
To define these thresholds we use the so called SURE thresholding technique
introduced by Donoho \& Johnstone 1995. This technique is based 
on finding an estimator of the signal that will minimize the expected
loss or risk defined as the mean value of 
$(1/N_p)\sum_{i=1}^{N_p}(T_{d_i}-T_i)^2$, where $T_i$ is the temperature 
at pixel $i$ in the original signal map and $T_{d_i}$ is the
estimator at pixel $i$ (temperature in the final denoised map). 
The minimization is finally achieved in the wavelet domain by choosing
a threshold value that minimizes the risk at each wavelet level (see 
for instance Ogden 1997).

Results of the errors in the map reconstruction are shown in Table 1. 
The map error is defined as:
\begin{equation}
\Bigl({{\sum_{i=1}^{n_{pixels}} (T_i-Td_i)^2}\over 
{\sum_{i=1}^{n_{pixels}} T_i^2}}\Bigr)^{1/2}\; .
\end{equation}
\noindent 
Performing 20 simulations (proved to be enough as results reached stable values) 
at each S/N level we have also calculated the
$1\sigma $ error.
The error improvement achieved with the denoising technique applied goes from 
factors of 3 to 5 for $S/N=3$ to $S/N=0.7$.

\begin{table}
  \begin{center}
  \caption[]{Reconstruction errors vs S/N.}
  \label{tab1}
  \begin{tabular}{c|c}\hline
  S/N & $\%$ map error $\pm 1\sigma$\cr
  \hline\hline

	0.7 & 26.3$\pm$0.4 \cr
	1.0 & 20.7$\pm$0.4 \cr
	2.0 & 13.3$\pm$0.2 \cr
	2.0 (n.u.) & 14.3$\pm$0.3 \cr
	3.0 & 10.3$\pm$0.2 \cr

  \hline\hline
  \end{tabular}
  \end{center}
\end{table}

It is also interesting to see how well the denoising method performs to 
reconstruct the temperature power spectrum. 
Mean values and $1\sigma $ errors of
the relative errors, $| \Delta C_{\ell}/C_{\ell}|$, are shown in Figure 3 for
three $S/N$ ratios and the case of non-uniform noise considered in this work. 
The $C_{\ell}$'s are reconstructed from the denoised 
maps with $|\Delta C_{\ell}/C_{\ell}| \le 10\%$ up to $l\sim 1000$ in cases $S/N\ge 1$.
This error can only be achieved 
up to an $l\sim 700$ in the $S/N=0.7$ case.
Higher order multipoles ($\ell \le 1500$) are reconstructed with $|\Delta C_{\ell}/C_{\ell}| \le 20\%$. 
Absolute relative errors and reconstructed $C_{\ell}$'s 
for a given map are presented in Figures 4 and 5 respectively. 

In order to check the performance of the SURE thresholding 
technique, knowing the original maps we can find the optimal 
threshold to get a reconstructed map with a minimum error (as defined
above).
In $S/N=1$ maps the optimal threshold is found to be $0.6-0.7 \sigma_n$. 
Thresholds between $0.3-1 \sigma_n$ do not make
substantial changes in the reconstructed map (see Table 2). The 
hard case included in that table stands for a case where all
coefficients below a signal-to-noise dispersion ration $<1.5$ are removed, 
leaving the others untouched.
For comparison, the error obtained comparing the signal plus noise map
with the original signal map is also presented in Table 2. We can see that
the error reconstruction achieved with the SURE technique equals
the one obtained with the optimal threshold.

\begin{table}
  \begin{center}
  \caption[]{Reconstruction errors vs threshold, S/N=1.}
  \label{tab2}
  \begin{tabular}{c|c}\hline
  Threshold & $\%$ map error \cr
  \hline\hline

	hard	& 23.5 \cr
	1.5 $\sigma_n$	& 21.7 \cr
	1.0 $\sigma_n$	& 20.7 \cr
	0.7 $\sigma_n$	& 20.5 \cr
	0.6 $\sigma_n$	& 20.6 \cr
	0.5 $\sigma_n$	& 20.7 \cr
	0.4 $\sigma_n$	& 21.0 \cr
	0.3 $\sigma_n$	& 21.3 \cr
\hline
	signal+noise    & 100.0\cr 
  \hline\hline
  \end{tabular}
  \end{center}
\end{table}

A comparison of wavelet techniques with Wiener filter (see for instance Press et al. 1994) has also been performed.
In relation to map reconstruction the error affecting the Wiener reconstructed
maps is comparable to the error for the
wavelet reconstructed maps, in all cases. 
However, in order to apply Wiener filter previous knowledge of signal power
spectrum is requiered.
Reconstructed and residual 
maps using both,
wavelets and Wiener filter, are shown in Figure 6.
Regarding the $C_{\ell}$s, performance of
Wiener filter is clearly worse than Wavelets for $\ell=1000-1500$, as can be
seen in Figure 4. For example, for a $S/N=1$ the $C_{\ell}$s are recovered 
using Wiener filter with
an error between 20$\%$ and 70$\%$ for $\ell$s between 1000 
and 1500 being this error smaller by a factor of 2-4 for 
Wavelet reconstruction. The error is also clearly larger for Wiener
reconstruction than for Wavelet reconstruction, up to $\ell\sim 2000$ in
cases with $S/N>1$.

We have checked for non-Gaussian features possibly introduced by the
non-linearity of the soft thresholding used in the wavelet methods applied
for denoising. 
Distributions of Skewness and Kurtosis have been
obtained for the original signal maps as well as for the denoised
ones. No significant differences can be appreciated between both
distributions. However this method could not be good enough to 
detect non-Gaussian features. 
As recently claimed by Hobson et al. 
(1998), the analysis of the distribution of wavelet coefficients is one of the most efficient
methods to detect them. We have performed a similar analysis using the
Daubechies 4 multiresolution wavelet coefficients. These coefficients are
Gaussian distributed in the case of a temperature Gaussian random field.  
The application of soft thresholding to the wavelet coefficients at a 
certain level clearly changes the Gaussian distribution by removing all 
coefficients whose absolute values are below the imposed threshold and 
shifting the remaining ones by that threshold. As an example, in
the previous case $S/N=1$ the kurtosis 
of the diagonal level 3 distribution changes from $3.3\pm 0.1$ to
$34\pm 10$ ! (notice that the change strongly 
depends on the threshold imposed). This result is not surprising as
any non-linear method used for denoising or foreground separation 
will introduce non-Gaussinity at different levels in the reconstructed map.
Fortunately there are two ways of overcoming the question of determining
the Gaussianity of the CMB signal. One way would be to check the Gaussian
character of the data before applying denoising to maps affected by
Gaussian noise. We have checked this
by looking at the multiresolution wavelet coefficient distributions in the
case of $S/N=1$. The addition of white noise didn't change the 
mean value and error bar of the kurtosis. 
The second way would be applying
a linear denoising method. 
We have used a simple one consisting in removing all detail coefficients at 
levels
with signal-to-noise dispersion ratio $<1.5$
(notice that $1.5$ corresponds the upper value of the threshold interval 
where soft thresholding was applied). This method is equivalent to applying 
hard thresholding with 
a threshold above all the coefficients. The errors of the reconstructed map 
and its 
corresponding $C_l$s 
increase slightly compared to the
SURE thresholding method (see table 2and top-left panel of figure 4). 
The same hard thresholding linear method will
give even better results using 2-D Wavelets with two scales of
dilation (Sanz et al. 1999) instead of the one-scale multiresolution
techniques, since the former works with many more
resolution levels being therefore more selective in removing the 
coefficients.    



\setcounter{figure}{5}
\begin{figure}
 \epsfxsize=6in
 \caption{
$12^{\circ}.8\times 12^{\circ }.8$ maps
of the cosmological signal (top left), signal plus noise with $S/N=1$
(top right),
denoised map using a soft thresholding as explained in the text 
(middle left) and
residual map obtained from the CMB signal map minus the denoised one (middle
right). For comparison a denoised map using Wiener filter
is presented in the bottom left panel together with 
the residuals in the bottom right panel. 
}
 \label{f6}
\end{figure}


\section{Discussion and Conclusions}

A wavelet multiresolution technique has been presented and used to analyse
and denoise CMB maps. This method has been proved to be one of the best 
to reconstruct observed CMB maps as well as power
spectra by removing a significant percentage of the noise. The analysis has 
been carried out assuming a uniform Gaussian noise as would be expected in a 
small sky patch, e.g. $12^{\circ}.8\times 12^{\circ}.8$,  
observed by satellite scans. Analysis of whole sky 
CMB maps using wavelets will be performed in a future work. 
Since these data will be affected by non-uniform noise, 
the use of wavelet techniques to localize map features will be highly 
suitable. 


A semi-analytical calculation of the variance of the
wavelet coefficients has been presented. The behaviour of the variance
of the {\it detail} coefficients
is given for a standard CDM model in the case of Haar and Mexican Hat
bases. The acoustic peaks can be noticed in the wavelet coefficient 
variance represented in Figure 1. Moreover, these peaks
are better defined for the Mexican Hat wavelet system since these wavelets are
more localized than the Haar ones.

Denoising of CMB maps has been carried out 
by using a 
signal-independent prescription, the SURE thresholding method.
The results are
model independent depending only on the observed data. 
However, a good knowledge of the noise affecting
the observed CMB maps is required. For a typical case of $S/N\sim 1$
the high order {\it detail} coefficients are dominated by the signal, whereas
the lowest ones are noise dominated. This behaviour is due to 
the expected dependence of the temperature power spectrum, 
$C_{\ell} \propto l^{-2}$. 
The applied wavelet method is able to reconstruct maps with 
an error improvement factor between 3 and 5 and the CMB power
spectrum of the denoised maps carries relative errors below $20\%$ up to 
$l\sim 1500$ for $S/N \ge 1$. 
We have also checked that SURE thresholding
methods are providing thresholds in agreement with the optimal ones.

For comparison Wiener filter has also been applied to the simulations considered in this paper. This method reconstructs CMB maps after denoising
with errors comparable to the Wavelet method we propose,
as shown in Figure 6.
However, the $Cl$s of the denoised maps obtained applying Wiener filter
have relative errors larger than a factor of 2 than the relative errors
of the $C_{\ell}$s obtained from the wavelet reconstructed maps 
in the range $\ell=1000-1700$.
In addition we have applied a Maximum Entropy Method (MEM) to the maps used in 
this work, with the
definition of entropy given by Hobson \& Lasenby (1998). 
This method provides reconstruction errors 
at the same level
as multiresolution wavelet methods. However, 
the later are easier (not requiring iterative processes) 
and faster ($0(N)$) to apply than MEM.

A possible handicap of denoising methods based on soft thresholding of
wavelet coefficients as well as other non-linear methods are the
non-Gaussian features introduced in the reconstructed map. However one
can still detect the possible intrinsic non-Gaussianity 
of the CMB signal by studying it in the signal plus noise map using
the wavelet coefficient distribution.
Moreover
a valid reconstruction 
can be obtained
by applying a ``hard'' thresholding linear method as discussed in the
text.


In a different work, we are studying the case of using a wavelet method
based on two scales of dilation (Sanz et al. 1999). Though this method has the 
advantage of keeping information on two different scales,
for the purpose of denoising 
both methods give comparable results. 
The linear hard thresholding method is expected to
perform better for 2-D wavelets than for 
multiresolution ones as the former works with many more resolution levels.

Summarizing, the main advantages of the wavelet method are:
to provide local information of the contribution from different scales,
to be computationally very fast $0(N)$, absence of tunning parameters and the most
important, the good performance on denoising CMB maps. The best reconstruction
is achieved using soft thresholding techniques. Concerning the Gaussianity of
the signal one can apply the suggested linear method for denoising. 
Moreover, the soft thresholding technique will provide a good reference map
and power spectrum for the signal, that can be used to check the quality 
of other reconstructions based on linear methods.  
Wavelets are also expected to be a very
valuable tool to analyse future CMB maps as those that
will be provided by future missions like MAP and Planck.



\section*{Acknowledgments}

We acknowledge 
helpful discussions with B. Dugnol, C. Fern\'andez, 
J.M. Gut\'\i errez, A.W. Jones and L. Tenorio. 
This work has been supported by the Spanish DGES Project
no. PB95-1132-C02-02, Spanish CICYT Acci\'on Especial no. ESP98-1545-E.
JLS, LC, EMG and RBB acknowledge finantial support from the USA-Spain 
Science and Technology Program, ref. 98138.  
RBB has been
supported by a Spanish MEC fellowship. LT acknowledges partial 
financial support from the Italian ASI and CNR.


\end{document}